\documentstyle[11pt,aaspp4]{article}

\received{1999 April }
\accepted{ }

\slugcomment{To appear in {\it Pub. of the Astr. Soc. of the Pac.},
August, 1999} 

\lefthead{Filippenko et al.}
\righthead{Black Hole in Nova Velorum 1993}

\begin{document}

\title{A Black Hole in the X-Ray Nova Velorum 1993}

\vspace{2cm}

\author{Alexei V. Filippenko, Douglas C. Leonard, Thomas Matheson,
Weidong Li, Edward C. Moran\altaffilmark{1}, and Adam G. Riess}
\affil{Department of Astronomy, University of California, Berkeley, 
California  94720-3411}
\affil{Electronic mail: (alex, dleonard, tmatheson, wli, emoran, 
ariess)@astro.berkeley.edu}

\vspace{2cm}



\altaffiltext{1}{Chandra Fellow.}


\begin{abstract}

We have obtained 17 moderate-resolution ($\sim 2.5$~\AA) optical spectra of the
Galactic X-ray Nova Velorum 1993 in quiescence with the Keck-II
telescope. Cross-correlation with the spectra of late-type dwarfs (especially
K7--M0) yields excellent radial velocities for the secondary star. The orbital
period ($P$) is $0.285206 \pm 0.0000014$~d, and the semiamplitude ($K_2$) is
$475.4 \pm 5.9$ km s$^{-1}$. Our derived mass function, $f(M_1) = PK_2^3 /2\pi
G = 3.17 \pm 0.12~M_\odot$, is close to the conventional absolute limiting mass
for a neutron star ($\sim$ 3.0--3.2~$M_\odot$) --- but if the orbital
inclination $i \lesssim 80^\circ$ (given the absences of eclipses), then $M_1
\gtrsim 4.2-4.4~M_\odot$ for nominal secondary-star masses of $0.5~M_\odot$
(M0) to $0.65~M_\odot$ (K6).  Even if the secondary is quite undermassive
(e.g., $M_2 = 0.3~M_\odot$), we derive $M_1 \gtrsim 3.9~M_\odot$.  The primary
star is therefore almost certainly a black hole rather than a neutron star.

  Fits to the wings of the double-peaked H$\alpha$ emission line yield
approximate radial velocities for the compact primary. The velocity curve has a
semiamplitude ($K_1$) of $65.3 \pm 7.0$ km s$^{-1}$, but with a phase offset by
$237^\circ$ (rather than $180^\circ$) from that of the secondary star.  Under
the assumption that the observed semiamplitude reflects the motion of the
primary (despite this offset, which is common), the mass ratio $q = M_2/M_1 =
K_1/K_2 = 0.137 \pm 0.015$, and hence for $M_2 $= 0.5--0.65~$M_\odot$ we derive
$M_1 =$ 3.64--4.74~$M_\odot$.  The constraints from $q$ and the mass function
yield $M_1 = 4.4~M_\odot$ and $i \approx 78^\circ$ if we use a normal K7--K8
secondary ($M_2 \approx 0.6~M_\odot$). Indeed, consistency cannot be achieved
for $M_2 \lesssim 0.59~M_\odot$ if the maximum inclination estimate
($80^\circ$) is correct. An adopted mass $M_1 \approx 4.4~M_\odot$ is
significantly below the typical value of $\sim 7~M_\odot$ found for black holes
in other low-mass X-ray binaries.

   Keck observations of MXB 1659--29 (V2134 Oph) in quiescence reveal a
probable optical counterpart at $R = 23.6 \pm 0.4$ mag. This object recently
went into a new X-ray and optical ($R \approx 18$ mag) outburst, after
about 21 years of inactivity.

\end{abstract}


\medskip



%

\section{INTRODUCTION}

   An important class of binary systems has been identified in which a low-mass
secondary (companion) star orbits a probable black hole; see White, Nagase, \&
Parmar (1995), van Paradijs \& McClintock (1995), and Tanaka \& Lewin (1995)
for extensive reviews. In all cases they were first observed in outburst as
``X-ray novae" (or ``soft X-ray transients"). Their X-ray spectra are generally
characterized by a prominent, ``soft" thermal component ($kT \approx 1$ keV) as
well as a ``hard" power-law tail extending to very high energies (0.1--1
MeV). During outburst, the radiation from these and other low-mass X-ray
binaries (LMXBs; the ``low mass" refers to the secondary star) is emitted
predominantly by the accretion disk surrounding the primary star.

   A lower limit to the mass of the primary ($M_1$) in a given X-ray transient
can be measured when it returns to quiescence. At that time, light from the
secondary star contributes significantly to (or even dominates) the visible
spectrum, and the secondary's radial-velocity curve can be determined with a
series of time-resolved spectra (see Cowley 1992 for a review). The orbital
period ($P$) and semiamplitude ($K_2$) of the secondary yield the mass function
of the primary, $f(M_1) = PK_2^3 /2\pi G = M_1^3\, {\rm sin}^3 i/(M_1 +
M_2)^2$, where $i$ is the inclination of the orbital plane to our line of
sight. Clearly, $f(M_1)$ provides an absolute lower limit to the mass of the
primary; only if $M_2 = 0$ and $i = 90^\circ$ is $M_1 = f(M_1)$. It is
generally acknowledged that if the primary is dark and has $f(M_1) \gtrsim
3.2~M_\odot$, it is probably a black hole, since the theoretical upper limit to
the mass of a normal neutron star is $\sim$ 3.0--3.2~$M_\odot$ (Friedman, Ipser,
\& Parker 1986; but see Friedman \& Ipser 1987, as well as Bahcall, Lynn, \&
Selipsky 1990, for conditions that might allow neutron stars to exceed this
nominal limit).

   The six best examples of LMXBs whose mass function exceeds (or is
close to) $\sim 3~M_{\odot}$, along with the derived values of $f(M_1)$ (in
units of $M_{\odot}$) and references, are as follows in chronological order of
the mass-function measurement: A0620--00 = V616 Mon ($3.18 \pm 0.16$,
McClintock \& Remillard 1986; $2.72 \pm 0.06$, Marsh, Robinson, \& Wood 1994;
$2.91 \pm 0.08$, Orosz et al. 1994), GS~1124--68 = Nova Mus 1991 ($3.1 \pm
0.4$, Remillard, McClintock, \& Bailyn 1992; $3.01 \pm 0.15$, Orosz et
al. 1996), GS~2023+338 = V404 Cyg ($6.26 \pm 0.31$, Casares, Charles, \& Naylor
1992; $6.08 \pm 0.06$, Casares \& Charles 1994), GRO J1655--40 = Nova Sco 1994
($3.16 \pm 0.15$, Bailyn et al. 1995; $3.24 \pm 0.09$, Orosz \& Bailyn 1997),
GS~2000+25 = Nova Vul 1988 ($5.02 \pm 0.47$, Casares, Charles, \& Marsh 1995;
$4.97 \pm 0.10$, Filippenko, Matheson, \& Barth 1995a, slightly revised to
$5.01 \pm 0.15$ [not $\pm 0.12$] by Harlaftis, Horne, \& Filippenko 1996), and
Nova Oph 1977 ($4.0 \pm 0.8$, Remillard et al. 1996; $4.86 \pm 0.13$,
Filippenko et al. 1997, slightly revised to $4.65 \pm 0.21$ by Harlaftis et
al. 1997).  Here we add a seventh object to this list: Nova Vel 1993 = GRS
1009--45, with a derived mass function of $3.17 \pm 0.12~M_\odot$.

  Nova Vel 1993 was discovered on 12 September 1993 with the WATCH all-sky
monitor aboard {\it Granat} (Lapshov, Sazanov, \& Sunyaev 1993; Lapshov et
al. 1994) and with BATSE on the {\it Compton Gamma-Ray Observatory} (Harmon et
al. 1993). Its spectrum exhibited an ``ultrasoft" hump at low energies
($\lesssim 1$ keV) and a power-law tail out to at least 100 keV (Kaniovsky,
Borozdin, \& Sunyaev 1993), typical of X-ray binaries in which the compact
object is a black hole. Two months later (17 November), Della Valle \& Benetti
(1993) discovered a blue optical counterpart at $V \approx 14.6$ mag, but
reasonable estimates suggest that the magnitude at the time of outburst was $V
= 13.8 \pm 0.3$ (Della Valle et al. 1997). Optical photometry conducted by
Bailyn \& Orosz (1995) about half a year after the primary outburst showed the
presence of a secondary outburst and several mini-outbursts, again reminiscent
of black-hole X-ray novae. Della Valle et al. (1997) suggested an orbital
period of about 4 hours, but a more reliable period of $6.86 \pm 0.12$ hours
was obtained by Shahbaz et al. (1996).  The spectral type of the secondary star
in the binary system was estimated to be late-G/early-K by Shahbaz et
al. (1996), and later than G5--K0 by Della Valle et al. (1997).

   On 1998 January 25 (UT dates are used throughout this paper), we obtained
several $R$-band images of the field of Nova Vel 1993 with the Low Resolution
Imaging Spectrometer (LRIS; Oke et al.  1995) at the Cassegrain focus of the
Keck-II telescope. As can be seen in Figure 1, which shows a subset of one
image (seeing $0.65''$), the nova was in quiescence by this time, with $R =
21.2 \pm 0.2$ mag.\footnote[2]{We adopt the magnitudes of comparison stars
quoted in Table 1 of Della Valle et al. (1997). There appears to be a numbering
mismatch, or errors in the photometry, of some stars in Table 1 of Shahbaz et
al. (1996): for example, their Stars 2, 5, and 7 should have comparable
magnitudes, yet they are listed as being very different.}  To determine whether
Nova Vel 1993 should be considered a good dynamical black-hole candidate, we
decided to obtain a radial-velocity curve with LRIS. Our group had already
obtained excellent results in this way for the black-hole candidates GRO
J0422+32 (Filippenko, Matheson, \& Ho 1995b; Harlaftis et al. 1999), GS~2000+25
(Filippenko et al. 1995a; Harlaftis et al. 1996), and Nova Oph 1977 (Filippenko
et al. 1997; Harlaftis et al. 1997), all of which are comparably faint.

\section{OBSERVATIONS AND REDUCTIONS} 

   Nova Vel 1993 was observed with LRIS in 1998 during the nights of January
25, February 1, March 5--6, and May 2, as well as in 1999 during the night of
January 21. A journal of useful observations is given in Table~1.  (The
spectra obtained on 1998 May 2, and a few spectra on other nights, were of
marginal quality and are not considered here.)  Given the object's far
southerly declination ($-45^\circ$), and the restrictive southwest azimuth
limit on Keck-II ($\lesssim 185^\circ$), observing was restricted to $\sim 1.5$
hours per night; hence, a number of different observing runs separated by a
range of intervals was needed to avoid serious aliasing of the orbital period.
Conditions were always clear, and the seeing was about $1.2^{\prime\prime}$,
reasonably good considering the high airmass ($\sim 2.4$). Typical exposure
times were 1100~s.

   The long slit of width $1^{\prime\prime}$ was oriented at a position angle
(PA) of $160^\circ$ for all observations except that of 1998 January 25.  This
was close to the parallactic angle at the time of observation, thereby reducing
differences in the relative amount of light lost at different wavelengths
(Filippenko 1982). At such a PA, the slit went directly through Star A, which
is much brighter than the quiescent nova and only $\sim 1.6^{\prime\prime}$ SE
of it.  (Della Valle et al. 1997 incorrectly state that Star A is SW of the
nova.)  However, since the direction of atmospheric dispersion (i.e., the
parallactic angle) at the time of observation coincided with the nova/Star-A
orientation, the degree of contamination from Star A was nearly independent of
wavelength. The slit also partially intersected Stars B and C (Fig. 1), but
their light did not contaminate that of the nova; see Figure 2, which plots the
intensity of light along the slit at the wavelength of H$\alpha$ (thereby
accentuating the nova's contribution).

   We used a Tektronix $2048 \times 2048$ pixel CCD with a scale of
$0.215^{\prime\prime}$ pixel$^{-1}$ ($0.43^{\prime\prime}$ per binned pixel in
the spatial direction).  The 1200 grooves mm$^{-1}$ grating, blazed at
7500~\AA, resulted in a wavelength range of $\sim$ 5650--6950~\AA, and a
full-width at half-maximum (FWHM) spectral resolution of $\sim 2.5$~\AA\ ($\sim
120$ km s$^{-1}$) with the $1^{\prime\prime}$ slit, essentially identical to
what we used in our previous studies of black-hole X-ray novae. Thirteen
velocity standards with spectral types in the range G5~V--M2~V were observed
with the same setup, given the estimated classification of the secondary star
in Nova Vel 1993 (Shahbaz et al. 1996; Della Valle et al. 1997).  Also, spectra
of some sdF stars (Oke \& Gunn 1983) were obtained for flux calibration and
removal of telluric absorption lines.

   Cosmic rays were eliminated from the two-dimensional spectra through
comparison of pairs of consecutive exposures. The two-dimensional spectra were
bias-subtracted and flattened in the usual manner. The wavelength scale was
determined from polynomial fits to the positions of emission lines in spectra
of Hg-Ne lamps obtained with the telescope at (or near) the position of each
object.  To ensure accurate wavelength calibration, final corrections ($\sim
0.0$--0.3~\AA) to the wavelength solution were obtained from night-sky emission
lines in the spectra of the nova.

   We used the APALL task in IRAF\footnote[3]{IRAF is distributed by the
National Optical Astronomy Observatories, which are operated by the Association
of Universities for Research in Astronomy, Inc., under cooperative agreement
with the National Science Foundation.} to optimally extract (Horne 1986)
one-dimensional, sky-subtracted spectra of Nova Vel 1993. However, to minimize
contamination from Star A, we extracted only a few rows of the CCD centered on
the position of the nova; as shown in Figure 2, these were rows 3--5 from the
center of Star A, in the left wing of its spatial profile, corresponding to a
displacement of 1.3--2.1$^{\prime\prime}$.  To further remove contamination, we
also extracted the same rows (3--5) in the right wing of Star A's spatial
profile, and subtracted this spectrum from that of the nova.\footnote[4]{This
extraction determined the flux scale of the subtracted spectrum, but its
signal-to-noise (S/N) ratio was improved by using an extraction of the entire
right half (rows 0--5) of Star A's spatial profile and scaling to the spectrum
of rows 3--5.}  We found that Star A typically contributed 40\% (65\% on
January 21) of the flux in the original extraction of Nova Vel 1993. In general
(but, surprisingly, not on January 21), the subsequent analysis worked best on
these fully decontaminated spectra of the nova, whose typical S/N ratio per
final 0.75~\AA\ bin is $3.3 \pm 0.4$ in the continuum.

\section{RESULTS}

\subsection{Unphased Average Spectra}

   Two unphased average spectra of Nova Vel 1993 are shown in Figure 3, with
that of 6 March 1998 offset by 2.5 units. In the region of overlap, these
resemble the spectra shown by Shahbaz et al. (1996). The strongest emission
line is H$\alpha$, and there may be weak He~I $\lambda$5876 on 6 March, though
the latter line partially coincides with the Na~I~D absorption. Despite being
in quiescence, Nova Vel 1993 still exhibits variability; the equivalent width
(EW) of the H$\alpha$ emission line, which is unaffected by errors in flux
calibration, was somewhat larger on March 6 (76~\AA) than on January 21
(60~\AA). Note that this emission is considerably weaker than in GRO J0422+32
(EW $\approx 250$~\AA; Filippenko et al. 1995b), but comparable to that in
GS~2000+25 (EW $\approx 40$~\AA; Filippenko et al. 1995a) and Nova Oph 1977 (EW
= 25--85~\AA; Filippenko et al. 1997). As in the previous objects we have
studied, the H$\alpha$ line has two peaks ($\Delta v \approx 1200$ km
s$^{-1}$), more obvious in March than in January.

\subsection{Cross-Correlations}

   Following the procedure discussed in Filippenko et al. (1995b), we employed
the FXCOR package (``Release 9/13/93") in IRAF to cross-correlate the spectra
of Nova Vel 1993 with the 13 velocity standards (shown in Fig. 4, along with
two others from our previous studies).  The correlation was done over the
ranges 5980--6270~\AA\ and 6320--6500~\AA\ to avoid the H$\alpha$ and He~I
$\lambda$5876 emission lines, Na~I~D absorption, the 6270~\AA\ interstellar
line, and poorly subtracted [O~I] $\lambda$6300 night-sky emission. In almost
all cases a definitive correlation peak was obvious. Typical values of the
Tonry \& Davis (1979) significance threshold were quite high, $R \gtrsim 3$,
with a few as high as $\sim 6$. As in our previous studies, we adopted the
FXCOR $1\sigma$ uncertainties reduced by a factor of 2.77; the Fourier
transform properties of Gaussians (the functions used in fitting the
cross-correlation peaks) were used to determine that FXCOR overestimates the
Tonry \& Davis uncertainties by this amount.

  The strongest formal correlation was obtained with BD+00~3090, which is
officially listed as an M0~V star (Upgren et al. 1972), but over our spectral
range it also looks very similar to K7~V and K8~V stars; see Figure 4. Indeed,
the correlation was insignificantly lower with K7~V and K8~V stars, but far
inferior with K5~V and M1~V stars. (We did not observe K6~V and K9~V stars.)
Thus, we conclude that the secondary star lies somewhere in the range K7~V
through M0~V, and possibly as early as K6~V. This is a slightly later spectral
type than preferred by Shahbaz et al. (1996; late-G/early-K) and Della Valle et
al. (1997; later than G5--K0), though the former authors note that their
derived mean density for the secondary (2.4 g cm$^{-3}$) suggests a K5~V
star. Our phased-average spectrum of the secondary star (see below) supports
the late-K classification.

   Using the radial velocities evaluated from the correlations with BD+00~3090
(corrected for the radial velocity of BD+00~3090 itself; 46.6 km s$^{-1}$,
Evans 1967), we conducted a non-linear least-squares fit (i.e., a $\chi^2$ fit;
Press et al. 1986, p. 521) to obtain the best cosine curve to match the data
(Fig. 5). The four-parameter fit (zero point, semiamplitude, period, and phase)
yielded a systemic velocity of $\gamma_2 = 40.7 \pm 4.0$ km s$^{-1}$, a
semiamplitude of $K_2 = 475.4 \pm 5.9$ km s$^{-1}$, a period of $P = 0.285206
\pm 0.0000014$~d (6.84~hr), and a starting time (heliocentric Julian day) for
the phase of $T_0 =$ HJD 2,450,835.0661 $\pm$ 0.0007, where $T_0$ is defined as
the point of maximum {\it redshifted} velocity.  All the uncertainties are the
formal $1\sigma$ values derived from the $\chi^2$ fit, but that of $T_0$ may be
an underestimate because the choice of the range in which to search for $T_0$
is unconstrained by the data. A better measurement of the systemic velocity is
$\gamma_2 = 30.1 \pm 5.0$ km s$^{-1}$, the weighted average of values obtained
with the nine best standard stars.  Note that when Star A (Fig. 1) was
cross-correlated with the velocity standards, its derived velocities were
constant to within $\pm 7$ km s$^{-1}$ (full range); thus, our results for Nova
Vel 1993 are not an artifact of telescope position or instrument orientation,
and our removal of the contamination by Star A (spectral type $\sim$K1) was
effective.

   It is interesting to examine the distribution of possible periods resulting
from our series of observations. In Figure 6, which shows ${\chi}^2$ versus
trial period, there are several major groups of possible periods; the structure
within each group results from ambiguities in the counting of cycles between
widely separated epochs of observation (for example, from January through March
1998). Until we obtained the observations of 1999 January 21, the periods near
0.22~d and 0.285~d were equally probable, with those near 0.18~d and 0.4~d
distinctly inferior. With the additional data, however, we are now reasonably
confident that 0.285~d (6.84~hr) is correct. Based on observations of
photometric modulation observed during the decline from outburst, Bailyn \&
Orosz (1995) had speculated that the orbital period might be 1.6~hr or $\sim
3$~d, but these are clearly excluded by our radial velocity measurements.  The
possible presence of ``superhumps" in the optical light curve obtained 4 months
after the primary outburst led Della Valle et al. (1997) to deduce an orbital
period of 4~hr, closer yet still incorrect. Our period is essentially identical
that found by Shahbaz et al. (1996; $6.86 \pm 0.12$~hr) from Gunn $R$-band
ellipsoidal modulations in quiescence.  Note that the secondary in Nova Vel
1993 must be a dwarf; with an orbital period of only 6.84~hr, the compact
primary would be inside a giant or subgiant.

   The formal reduced ${\chi}^2$ (${\chi}_{\nu}^2$) for the best fit is 1.55
(17 points, 13 degrees of freedom). Given that the velocity uncertainties given
by the Tonry \& Davis (1979) method don't reflect external errors such as
miscentering of the object in the slit, they could easily be too small. In our
case, increasing them by only 25\% yields a reduced ${\chi}_{\nu}^2$ of 1.0.
Moreover, had we assigned uncertainties to the adopted times (i.e., phases) of
the observations, and performed the fit to simultaneously minimize residuals in
both velocity and phase, we would have obtained a smaller value of
${\chi}_{\nu}^2$. (The effective times of the observations can differ from the
calculated midpoints due to variations in observing conditions --- seeing,
transparency, etc.)

\subsection{The Phased Average Spectrum}

  Having determined the orbital parameters of the secondary star, we obtained
its master ``rest-frame" spectrum by averaging the 17 spectra after Doppler
shifting each one to zero velocity.  This represents a total integration time
of $\sim$5~hr, but of course the S/N ratio is lower than for a single 5-hr
exposure (ignoring cosmic rays) due to the increase in readout noise. Note that
the H$\alpha$ and He~I emission lines are smeared out by this process, since
they are produced in the accretion disk around the compact primary star.
Similarly, interstellar absorption lines become less distinct.

   Figure 7 shows the phased average spectrum of Nova Vel 1993, in comparison
with its typical spectrum (obtained at 11:14 UT on 1998 Feb. 1), the unphased
average around H$\alpha$, and the spectrum of the M0~V velocity standard
star. Scrutiny of the phased average spectrum reveals stellar absorption lines
that are also present in the M0~V star (e.g., the Ca + Fe blend at 6498~\AA),
but those of Nova Vel 1993 are weak. While this could indicate that the
spectral type of the secondary star is substantially earlier than M0, tests
show that several other factors are likely to be more important: (1) rotational
broadening, perhaps 50--100 km s$^{-1}$ (e.g., Wade \& Horne 1988; Harlaftis et
al. 1996); (2) orbital broadening, typically $4K_2 T/P = 80$--90 km s$^{-1}$
(Filippenko et al. 1995b); and, most importantly, (3) contamination by the
featureless continuum of an accretion disk.

   We attempted to quantify the contribution of the accretion disk in Nova Vel
1993 by comparing its phased average spectrum with that of the M0~V star,
broadening and diluting the latter by various amounts.  A good match to the
depths of the narrow absorption lines was found with accretion disk
contamination fractions of 60\%--70\% of the total flux density at $\sim
6300$~\AA. Because of the uncertainties in the adopted parameters (broadening,
spectral type) and the relatively low S/N ratio of the Nova Vel 1993 spectrum,
we cannot confidently exclude contributions somewhat outside this range, but it
is clear that the accretion disk dominates the spectrum even in quiescence.

   No absorption line of Li~I $\lambda$6708 is visible in our phased-average
spectrum, to a $3\sigma$ upper limit of $\sim 0.1$~\AA.  An unexpectedly strong
Li~I line (EW = 0.25--0.48~\AA) is seen in the spectra of several X-ray novae
(Mart\'\i n et al. 1994, and references therein; Filippenko et al. 1995a), but
not in others (such as Nova Oph 1977; Filippenko et al. 1997; Harlaftis et
al. 1997).

\subsection{H$\alpha$ Measurements}
   
    To investigate the motion of the compact primary in Nova Vel 1993, we fit a
Gaussian to the high-velocity wings ($v \approx$ 800--2000 km s$^{-1}$) of
the H$\alpha$ line in each individual spectrum using the IRAF task SPECFIT.
The regions of the fit (including the continuum) were 6400--6545~\AA\ and
6580--6710~\AA; the double-horned core of the line
(Fig. 3) was excluded. The velocity of the Gaussian peak and its formal
$1\sigma$ uncertainty were adopted for each spectrum (Table 1).

   As shown in Figure 8, the derived velocity tends to be low in the first half
of the orbit and high in the second half, perhaps suggesting periodic
behavior. We used a least-squares fit to determine the H$\alpha$
radial-velocity cosine curve, forcing the data to have the period found for the
secondary star (0.285206~d) but allowing all other parameters to vary. The {\it
formal} results (Fig. 8; ${\chi}_{\nu}^2 = 1.34$) are as follows: $\gamma_1 =
4.6 \pm 6.2$ km s$^{-1}$, $K_1 = 65.3 \pm 7.0$ km s$^{-1}$, and a zero point in
the phase of HJD $2,450,834.9686 \pm 0.0093$~d. The value of $\gamma_1$ is
inconsistent with that of $\gamma_2$, but this is often the case in
X-ray novae (e.g., Nova Oph 1977, Filippenko et al. 1997).

   Once again, the zero point in the phase is the maximum {\it redshifted}
velocity; it implies that the compact object is 237$^\circ$ out of phase with
the companion star, rather than the expected 180$^\circ$. This is comparable to
the offsets in GRO J0422+32 ($253^\circ$; Filippenko et al. 1995b) and
GS~2000+25 ($260^\circ$; Filippenko et al. 1995a), as well as in A0620--00 and
Nova Mus 1991 (Orosz et al. 1994). To date, there is no satisfactory
quantitative explanation for these distortions, but they suggest that the
accretion disk often has a nonaxisymmetric distribution of surface brightness,
noncircular velocities, or a warp. They also cast some doubt on the use of
H$\alpha$ radial-velocity curves to determine the motion and mass of the
primary star. On the other hand, the mass ratios determined in this manner are
frequently quite consistent with those obtained with independent techniques
(e.g., A0620--00, Orosz et al. 1994; GS 2000+25, Harlaftis et al. 1996; Nova
Oph 1977, Harlaftis et al. 1997; GRO J0422+32, Harlaftis et al. 1999).  Thus,
here we will cautiously adopt the ratio of semiamplitudes as an estimate of the
mass ratio: $q = M_2/M_1 = K_1/K_2 = 0.137 \pm 0.015$.

\section{THE MASS OF THE COMPACT PRIMARY}

   From the semiamplitude ($K_2 = 475.4 \pm 5.9$ km s$^{-1}$) and period ($P =
0.285206 \pm 0.00000138$~d) of the radial velocity curve of the secondary star,
we find a mass function $f(M_1) = PK_2^3 /2\pi G = 3.17 \pm 0.12~M_\odot$. This
corresponds to the absolute minimum mass of the compact primary, and it is
close to the maximum gravitational mass of a slowly rotating neutron star
(3.0--3.2~$M_\odot$; see Chitre \& Hartle 1976, and the discussion in
Filippenko et al. 1995b).

    No evidence for eclipses is seen in the data of Bailyn \& Orosz (1995),
Shahbaz et al. (1996), or Della Valle et al. (1997); similarly, we do not see
significant variations in the apparent brightness of the secondary star over
the orbital period.  Hence, it is likely that the orbital inclination $i
\lesssim 80^\circ$, and the relation $f(M_1) = M_1^3\, {\rm sin}^3 i/(M_1 +
M_2)^2$ then implies that $M_1 \gtrsim$ 4.2--4.4~$M_\odot$ for nominal
secondary-star masses of 0.5--0.65~$M_\odot$ (M0--K6~V; Allen 1976).  Even if
the secondary is quite undermassive (e.g., $M_2 = 0.3~M_\odot$), as in some
X-ray binaries (e.g., van den Heuvel 1983), we derive $M_1 \gtrsim
3.9~M_\odot$.  The primary star is therefore almost certainly a black hole
rather than a neutron star.

  Adopting the mass ratio derived from the measured semiamplitude of the radial
velocity curve of the primary ($q = M_2/M_1 = K_1/K_2 = 0.137 \pm 0.015$), we
find $M_1 =$ 3.64--4.74~$M_\odot$ if $M_2 =$ 0.5--0.65~$M_\odot$. The
constraints from $q$ and the mass function yield $M_1 = 4.4~M_\odot$ and $i
\approx 78^\circ$ if we use a normal K7--K8 secondary ($M_2 \approx
0.6~M_\odot$). Indeed, consistency cannot be achieved for $M_2 \lesssim
0.59~M_\odot$ if the maximum inclination estimate ($80^\circ$) is correct. We
conclude that the secondary star cannot be substantially undermassive, and that
the orbital inclination is probably rather large, almost making Nova Vel 1993
an eclipsing binary. Our suggested inclination is significantly higher than the
nominal value derived by Shahbaz et al.  (1996) from their observed $R$-band
ellipsoidal modulations ($i = 44^\circ \pm 7^\circ$). However, these authors
admit that when contamination by light from the accretion disk is included,
their allowed range for the inclination is $37^\circ$--$80^\circ$. Of course,
in view of the unexplained phase offset between the expected and observed
H$\alpha$ radial velocity curves (\S~3.4), it is also possible that our derived
value of $q$ is erroneous, thereby affecting our estimate of $i$.  Further
studies are needed to accurately determine the inclination.

   Recently, Bailyn et al. (1998) found that the distribution of masses of
putative black holes in LMXBs is very strongly peaked at $\sim 7~M_\odot$, only
V404 Cyg being a clear high-mass deviant. The number of objects in the sample
is still quite small, but if our mass estimate for Nova Vel 1993 is correct,
then it appears to be a low-mass counterexample ($M_1 \approx
4.4~M_\odot$). Another possible exception is GRO J0422+32 ($M_1 \approx
5~M_\odot$; Harlaftis et al. 1999).  An independent measure of the mass ratio
of Nova Vel 1993 from the rotational broadening of the secondary star's
absorption lines (e.g., Nova Oph 1977, Harlaftis et al. 1997; GRO J0422+32,
Harlaftis et al. 1999), together with better constraints on the inclination
derived from near-infrared ellipsoidal modulations (e.g., GS 2000+25, Callanan
et al. 1996), would provide a very useful check on our estimated mass for the
primary star.

   When we calculate the effective Roche lobe radius ($R_L$) of the companion
star from the relation of Paczy\' nski (1971; see also Eggleton 1983) and
Kepler's third law, we find that $R_L = 0.7~R_\odot$ if $M_1 \approx
4.4~M_\odot$ and $M_2 \approx 0.6~M_\odot$. This is only slightly larger than
the expected radius of a typical K8 dwarf ($R = 0.67~R_\odot$; Allen
1976). Thus, the secondary star may be just starting its evolution off the main
sequence.

\section{OBSERVATIONS OF MXB 1659--29 IN QUIESCENCE}

   As part of our effort to determine the mass functions of X-ray binaries, we
observed the field of MXB 1659--29. This burst source was discovered in 1976
October with SAS~3 (Lewin et al. 1976). It was initially considered unusual
because of its very stable burst intervals (Lewin 1977) and apparent absence of
constant emission, but such emission was found a year later (Lewin et al. 1978;
Share et al. 1978). An optical counterpart (now known as V2134 Oph) was
discovered by Doxsey et al. (1979) at $V = 18.3$ mag; it was quite blue, and
possibly exhibited emission lines of He~II $\lambda$4686 and C~III/N~III
$\lambda\lambda$4640--4650. Figure 2 of Doxsey et al. (in which N is up and E
to the left, although this isn't stated) shows $U$-band and $B$-band finder
charts for the object, from images taken on 1978 May 30 and June 1 with the
CTIO 4-m telescope.

   We obtained three dithered $R$-band images (exposure times of 60, 60 and
30~s) of the MXB 1659--29 field on 1999 February 9 with LRIS/Keck-II, at
airmass 1.8--1.9. These were bias-subtracted, flattened, registered, and
combined in the usual manner. A small subset of the resulting crowded image
(Galactic latitude 7.3$^\circ$) is shown in Figure 9a; the FWHM of stars is
measured to be $\sim 0.9''$.  Star A is close to the apparent position of V2134
Oph indicated in the relatively shallow charts published by Doxsey et
al. (1979). However, seven LRIS/Keck-II long-slit spectra (PA = 160$^\circ$) of
this star, each with a typical exposure time of 900--1000~s, do not reveal any
H$\alpha$ emission characteristic of accretion disks. Moreover,
cross-correlation of the individual spectra with the 13 velocity standards (G5
providing the best match) reveals no clear variability beyond the $\pm 15$ km
s$^{-1}$ level, and no systematic trend among consecutive exposures, casting
further doubt on Star A as the secondary.  H$\alpha$ emission is also weak or
absent in our noisy spectrum of Star B (Fig. 9a), and Star E has a spectral
type of M.
 
   Shortly before the completion of this paper, MXB 1659--29 went into outburst
again, after a hiatus of 21 years. During the interval 1999 April 2.06--3.47,
the Wide Field Camera on BeppoSAX detected a transient X-ray source coincident
with the position of MXB 1659--29 (in 't Zand et al. 1999). This object was
confirmed on April 5.83--6.05 with RXTE (Markwardt et al. 1999). Optical
observations (Augusteijn, Freyhammer, \& in 't Zand 1999) on April 3.41
revealed a bright new source ($V = 18.3 \pm 0.1$) at that location, with a
spectrum typical of LMXBs in an X-ray bright phase (emission lines of H~I,
He~II, C~III, and N~III). An image (exposure time 30~s) obtained on April 18
with LRIS/Keck-II is shown in Figure 9b (stellar FWHM = $0.65''$); the optical
counterpart ($R = 18.2 \pm 0.05$) is marked ``F."  This star is also visible at
the center of the circle in Figure 9a, barely at the detection limit.

   We measured the $R$ magnitude of V2134 Oph in quiescence (Fig. 9a) with the
technique of point-spread-function (PSF) fitting, where the PSF was determined
iteratively by subtracting faint stars near the ones chosen for the PSF.  The
zero point was obtained from twilight-sky observations of PG1525--071A,C
(Landolt 1992).  Star F, which we identify with V2134 Oph in quiescence, has $R
= 23.6 \pm 0.4$ mag. If Star F is just a chance superposition, then the true
optical counterpart is even fainter. Thus, any future attempts to obtain the
mass function of MXB 1659--29 (V2134 Oph) will be extremely difficult to
perform! As an aid for future photometry of this object, we note that the final
magnitudes for Stars A, B, C, D, and E are 19.6, 22.5, 23.1, 23.3, and 21.0,
respectively. The $1\sigma$ uncertainty is about 0.05 mag at the bright end
(primarily due to the dearth of photometric standards) and perhaps 0.2 mag for
stars at $R \approx 23$.

\section{CONCLUSIONS} 

  Our observations of Nova Vel 1993 provide a definitive mass function of $3.17
\pm 0.12~M_\odot$, and a likely mass of around $4.4~M_\odot$ for the primary
star. Thus, Nova Vel 1993 joins the small but growing list of secure Galactic
black holes first identified as X-ray novae.  However, its mass seems to be
lower than that of other objects in its class, bridging the apparent gap
between $\sim 3~M_\odot$ (the theoretical maximum mass of a neutron star,
though observed masses almost always yield $M = 1.0$--1.8~$M_\odot$; Thorsett
et al. 1993) and $\sim 7~M_\odot$ (the mass of most Galactic black holes in
binary systems with well-determined parameters).  It will be important to
measure the mass ratio and orbital inclination of the system with techniques
independent of the indirect ones used here, to confirm our estimates ($q =
0.137 \pm 0.015$; $i \approx 78^\circ$) and our derived mass.

   A similar study of MXB 1659--29 (V2134 Oph) eliminates several candidate
stars as the optical counterpart. The recent new outburst, which occurred just
prior to the submission of this paper, allows us to identify the quiescent nova
at $R = 23.6 \pm 0.4$ mag, unless this is an unrelated star superposed along
the line of sight.

\acknowledgments

  Data presented herein were obtained at the W. M. Keck Observatory, which
is operated as a scientific partnership among the California Institute of 
Technology, the University of California, and NASA. The Observatory was
made possible by the generous financial support of the W. M. Keck Foundation.
We thank T. Bida, R. Campbell, R. Goodrich, D. Lynn, G. Puniwai, H.  Rodriguez,
C. Sorensen, W. Wack, T. Williams, and other members of the Keck staff for
their able assistance. We are also grateful to N. Vogt, A. C. Phillips, and
D. C. Koo for obtaining a Keck image of MXB 1659--29 on April 18, after its
recent outburst. This work was supported by the NSF through Grant AST--9417213
to A.V.F.

\clearpage
\begin{deluxetable}{rccclrr}
\tablewidth{500pt}
\tablecaption{Journal of Observations and Radial Velocities \label{tbl-1}}
\tablehead{\colhead{HJD\tablenotemark{a}} &
\colhead{UT Date} &
\colhead{Exposure (s)} &
\colhead{Airmass\tablenotemark{b}} &
\colhead{Phase\tablenotemark{c}} & 
\colhead{$v_2$ (km s$^{-1}$)\tablenotemark{d}} &
\colhead{$v_1$ (km s$^{-1}$)\tablenotemark{e}} }

\startdata

838.96576 & 1998 Jan 25 &   900 & 2.53 & 0.67402  &   $-236.7 \pm 21.7$ & $123.0  \pm 62.3$  \nl
845.95655 & 1998 Feb 01 &  1200 & 2.46 & 0.18531  &   $219.4  \pm 18.9$ & $-36.10 \pm 21.7$ \nl
845.97085 & 1998 Feb 01 &  1200 & 2.39 & 0.23545  &   $52.09  \pm 17.1$ & $-21.29 \pm 25.1$  \nl
877.85456 & 1998 Mar 05 &   900 & 2.58 & 0.026917 &   $448.5  \pm 13.4$ & $-61.26 \pm 28.3$ \nl
877.86708 & 1998 Mar 05 &  1200 & 2.48 & 0.070815 &   $430.6  \pm 15.0$ & $-79.79 \pm 19.7$ \nl
877.88132 & 1998 Mar 05 &  1200 & 2.41 & 0.12074  &   $328.9  \pm 8.62$ & $-34.98 \pm 18.8$ \nl
877.89554 & 1998 Mar 05 &  1200 & 2.36 & 0.17060  &   $248.4  \pm 17.5$ & $-51.49 \pm 17.6$ \nl
877.90868 & 1998 Mar 05 &  1000 & 2.35 & 0.21667  &   $123.3  \pm 15.3$ & $-71.38 \pm 21.1$ \nl
877.92060 & 1998 Mar 05 &  1000 & 2.35 & 0.25847  &   $-61.19 \pm 15.5$ & $-86.64 \pm 21.0$ \nl
878.85228 & 1998 Mar 06 &  1100 & 2.58 & 0.52515  &   $-476.1 \pm 12.0$ & $75.89  \pm 25.1$ \nl
878.89168 & 1998 Mar 06 &  1100 & 2.37 & 0.66329  &   $-229.7 \pm 17.2$ & $63.71  \pm 22.3$ \nl
878.90475 & 1998 Mar 06 &  1100 & 2.35 & 0.70912  &   $-120.3 \pm 15.8$ & $40.81  \pm 20.7$ \nl
878.91785 & 1998 Mar 06 &  1100 & 2.36 & 0.75505  &   $ 11.61 \pm 21.8$ & $86.20  \pm 18.5$ \nl
1199.97900 & 1999 Jan 21 &  1200 & 2.52 & 0.46778  &  $-476.9 \pm 12.0$ & $14.13  \pm 41.8$ \nl
1200.00574 & 1999 Jan 21 &  1100 & 2.38 & 0.56154  &  $-451.1 \pm 8.61$ & $27.51  \pm 50.5$ \nl
1200.01881 & 1999 Jan 21 &  1100 & 2.35 & 0.60736  &  $-370.1 \pm 12.8$ & $31.60  \pm 66.0$  \nl
1200.03255 & 1999 Jan 21 &  1200 & 2.35 & 0.65554  &  $-272.4 \pm 27.6$ & $14.54  \pm 63.6$ \nl
\enddata
\tablenotetext{a}{HJD--2,450,000 at midpoint of exposure.}
\tablenotetext{b}{Airmass at midpoint of exposure.}
\tablenotetext{c}{Using $P = 0.285206$~d and $T_0 = 2,450,835.0661$.}
\tablenotetext{d}{Secondary star radial velocity.}
\tablenotetext{e}{H$\alpha$ centroid radial velocity, from fit to emission-line wings.}

\end{deluxetable}

\clearpage

\vfill
\eject

\begin{figure}
\caption[]{
$R$-band image of Nova Vel 1993, obtained on 1998 January 25. The
now-quiescent nova is about $1.7''$ NNW of Star A.}
\end{figure}

\begin{figure}
\caption[]{
Spatial cut of counts along the slit in a typical two-dimensional CCD spectrum
of Nova Vel 1993, computed from the average of 10 columns centered on the
wavelength of H$\alpha$. The nova is in the wing of Star A. Other stars
correspond to those shown in Fig. 1.}
\end{figure}

\begin{figure}
\caption[]{
Typical spectra of Nova Vel 1993, obtained on 1998 March 6 and 1999
January 21. Note the change in the shape and strength of the H$\alpha$
emission line. The top spectrum is offset by +2.5 units for clarity.}
\end{figure}

\begin{figure}
\caption[]{
Spectra of 15 radial velocity standards, arranged from early to late types.}
\end{figure}

\begin{figure}
\caption[]{
Radial-velocity curve of the secondary star in Nova Vel 1993,
derived from cross-correlations of individual spectra with those of the
M0~V star BD+00~3090. Two cycles are shown for clarity. 
The radial velocity of the comparison star, 
46.6 km s$^{-1}$, has been added to all values. The orbital period
is $0.285206 \pm 0.00000138$~d. Formal velocity error bars are $1\sigma$.}
\end{figure}

\begin{figure}
\caption[]{
Value of $\chi^2$ vs. trial period for Nova Vel 1993, using the 17
spectra discussed in this paper. Only the lowest values of $\chi^2$ are
shown; $\chi^2 \approx 10,000$ at typical periods near 0.35~d.}
\end{figure}

\begin{figure}
\caption[]{
Spectra of Nova Vel 1993 compared with a scaled spectrum of a
M0~V velocity standard. Constants have been added to the top three spectra
for clarity. The ``phased average"
spectrum in the rest frame of the companion star was obtained by combining
the 17 spectra after
Doppler shifting each one to zero velocity.
The narrow H$\alpha$ absorption in the unphased average spectrum {\it (top)} 
must be produced by gas associated with the accretion disk, not the
secondary star; otherwise, it would appear smeared out.}
\end{figure}

\begin{figure}
\caption[]{
Radial-velocity curve of the centroid of the H$\alpha$ emission
line, determined by fitting a Gaussian to the high-velocity wings. The
period was forced to be that of the secondary star, but all other
parameters were allowed to vary. Formal error bars are $1\sigma$.}
\end{figure}

\begin{figure}
\caption[]{
(a) $R$-band image of the field of MXB 1659--29 (V2134 Oph), obtained on 1999
February 9.  Various stars discussed in the text are marked. (b) $R$-band image
obtained on 1999 April 18. The nova in outburst is positionally coincident with
Star F in (a).}
\end{figure}

\end{document}